\documentstyle[epsfig,subfigure]{mn}
 at 10truept
\title [Redshift-space distortions of the PSCz galaxy catalogue]
{\vglue-3.0truecm
\centerline{\it\small For submission to Monthly Notices Letters}
\vglue 2.5truecm
    Application of Data Compression Methods to the
	 Redshift-space distortions of the PSCz galaxy catalogue
\author
     [A.N. Taylor et al.]
     {A.N. Taylor$^*$, W.E. Ballinger$^\dag$, A.F. Heavens$^*$, H. Tadros$^\dag$\\
     $^*$Institute for Astronomy,
     University of Edinburgh,
     Royal Observatory,
     Blackford Hill,
     Edinburgh,
     U.K.\\
     $^\dag$Astrophysics, University of Oxford, Keble Road, Oxford OX1 3RH\\
    	ant@roe.ac.uk, ballinger@astro.ox.ac.uk, afh@roe.ac.uk, 
	h.tadros1@physics.oxford.ac.uk}}
\def\bib{\parskip=0pt\par\noindent\hangindent\parindent
    \parskip =2ex plus .5ex minus .1ex}

\newcommand{\be}{\begin{equation}}
\newcommand{\ee}{\end{equation}}
\newcommand{\ba}{\begin{eqnarray}}
\newcommand{\ea}{\end{eqnarray}}

\newcommand{\nnb}{\begin{displaymath}}
\newcommand{\nne}{\end{displaymath}}

\newcommand{\s}{\mbox{\boldmath $s$}}

\newcommand{\C}{\mbox{\boldmath $C$}}
\newcommand{\D}{\mbox{\boldmath $D$}}

\newcommand{\Sb}{\mbox{\boldmath $S$}}

\newcommand{\Phib}{\mbox{\boldmath $\Phi$}}
\newcommand{\Vb}{\mbox{\boldmath $V$}}
\newcommand{\W}{\mbox{\boldmath $W$}}

\newcommand{\deltab}{\mbox{\boldmath $\delta$}}

\newcommand{\rhob}{\mbox{\boldmath $\rho$}}

\newcommand{\rhoo}{\mbox{\boldmath $\rho_0$}}
\newcommand{\eL}{{\cal L}}

\newcommand{\hMpc}{\,h^{-1}{\rm Mpc}}
\newcommand{\Mpch}{\,h{\rm Mpc}^{-1}}
\newcommand{\kms}{\,{\rm km}s^{-1}}

\newcommand{\rgl}{\rangle}
\newcommand{\lgl}{\langle}

\def\bib{\parskip=0pt\par\noindent\hangindent\parindent
    \parskip =2ex plus .5ex minus .1ex}

\begin{document}

\maketitle

\begin{abstract}
We apply a spherical harmonic analysis to the Point Source
Redshift Survey (PSCz), to compute the real-space galaxy power
spectrum and the degree of redshift distortion caused by peculiar
velocities.  We employ new parameter eigenvector and
hierarchical data compression techniques, allowing a much larger
number of harmonic modes to be included, and correspondingly
smaller error bars. Using 4644 harmonic modes, compressed to
2278, we find that the IRAS redshift-space distortion parameter
is $\beta = 0.39 \pm 0.12$ and the amplitude of galaxy clustering
on a scale of $k=0.1 \Mpch$ is $\Delta_{\rm gal}(0.1)=0.42 \pm 0.02$.
Combining these we find the amplitude of mass perturbations is
$\Delta_m(0.1)=(0.16\pm0.04) \Omega_m^{-0.6}$. A preliminary
model fitting analysis combining the PSCz amplitudes with the CMB
and abundance of clusters yields the cosmological matter density
parameter $\Omega_m=0.16\pm 0.03$, the amplitude of primordial
perturbations $Q=(8.4\pm 3.8) \times 10^{-5}$, and the IRAS bias
parameter $b=0.84\pm 0.28$.
\end{abstract}


\section{Introduction}

The extraction of cosmological parameters from surveys
has entered a new phase, with the advent of very large data sets.
But the prospect of accurately determining a wide range of parameters
is offset by the difficult task of manipulating these large
data sets without loosing important parameter information. In the
case of analysing near-all sky redshift surveys we have developed a
method based on the harmonic decomposition of the survey into
spherical harmonics and radial spherical Bessel functions (Heavens \&
Taylor 1995, Ballinger, Heavens \& Taylor 1995, Tadros et al 1999;
hereafter HT, BHT and T99, respectively).
The use of a spherical harmonic decomposition allows the accurate
analysis of the radial redshift-space distortion effect, without
using the small-angle or distant observer approximations, and
the natural inclusion of angular and radial window functions.
The parameters of interest are the degree of redshift-space
distortion, parameterised by\footnote{Since differently selected
galaxies cluster differently, each selection may have its own bias
parameter and $\beta$. In this paper we shall use $\beta$ and $b$ to refer
exclusively to IRAS selected galaxies.}
\be
    \beta \equiv \frac{f(\Omega_m)}{b}
\ee 
where $f(\Omega_m)=d \ln \delta /d \ln a \approx
\Omega^{0.6}_m$ (Peebles 1980) is the growth rate of perturbations
and $b$ is the linear bias parameter defined by \be
    P_{\rm gal}(k) = b^2 P_m(k),
\ee and an estimate of the undistorted galaxy power spectrum,
$P_{\rm gal}(k)$. This approach combines the spherical harmonic
decomposition with a mode-by-mode maximum likelihood analysis,
and has been applied to the 1.2Jy survey, a 1:6 subsample of the
IRAS galaxy survey (HT, BHT) and the full 1:1 sample, the PSCz
(T99). T99 used the most sophisticated analysis to date, and
found a distortion parameter of $\beta = 0.58 \pm 0.26$ (marginal
error) and an amplitude for the real space galaxy power spectrum
at $k=0.1 h {\rm Mpc}^{-1}$ of $\Delta_{0.1}=0.42\pm 0.03$, where $\Delta^2(k)=
k^3 P(k)/2 \pi^2$. 
In this paper we employ new data compression methods to
improve on this analysis and obtain more accurate parameter
determinations. We describe the methods more fully in Ballinger et al (2000).

The limiting factors in our previous analyses were CPU time and
stability.  To complete the likelihood analysis the data
covariance matrix of the full data set must be inverted at each
point in parameter space.  Small numerical errors can make this
procedure unstable; data compression (Section 2) is a great help
here, as the high signal-to-noise modes are well-behaved. The CPU
factor can also be an issue when one wishes to investigate
systematic effects in data sets. This is as much an issue in
analysis of the Cosmic Microwave Background (CMB) as it is in
galaxy redshift surveys, such as the PSCz, the 2-degree Field (2dF) or 
the Sloan Digital Sky Survey (SDSS).

The problem of analysing large data sets was addressed by Tegmark, Taylor
\& Heavens (1997; TTH) who considered the question of what was the optimal
transformation of the data for estimating a given parameter, where
the model data covariance matrix could be an arbitrary, nonlinear
function of the desired parameter. The optimal transformation should
have the properties of maximising the information content about the
parameter in the minimum number of eigenmodes. By ordering modes
the ones with the most information could be selected and the rest
discarded. That the data can be ordered this way can be understood
if one considers that the data may contain noisy or strongly correlated
modes that add little information about the parameter of interest.
Choosing the transformed data to have diagonal covariance also
decreases the computation time. While the data covariance is only
diagonal at one point in parameter space, the removal of correlated
and noisy data by trimming produces a numerically stable inversion
of the covariance matrix.

This procedure sounds similar to Principle Component Analysis
(PCA) or signal-to-noise eigenvalue analysis (SNA; Bond 1995,
Vogeley \& Szalay 1995), but has important differences.  We have
previously referred to the procedure as Karhunen-Lo\`{e}ve methods,
but it is in fact more general, so we shall refer to our method henceforth
as Optimal-Mode Analysis (OMA).  In this paper, we introduce 
two new methods for accurate parameter
estimation, and refer to the whole method as {\em Generalised Optimal Mode
Analysis}, or GOMA.  

The additional features of GOMA can be split into two parts, dealing with
multiparameter estimation and the stable handling of data compression. 
In multiparameter estimation it is useful to note that for
highly-correlated multiparameter estimates (highly-elongated likelihood
surfaces, not aligned with parameter axes), the marginal error in the
parameters is determined by the length of the longest likelihood
principal axis.  We therefore want to optimise to keep this length as short
as possible.  This process is called {\em parameter eigenmode
analysis}, and was introduced in Ballinger (1997), with some results
being presented in Taylor et al. (1997).  The second, optional part of GOMA
is to split the original data into subsets, optimising each subset,
and then combining the best modes together.  This procedure can be
used hierarchically, to reduce a very large number of modes to a
manageable size.  This process we refer to as {\em hierarchical data
compression}.  These methods will be detailed in a companion paper
(Ballinger et al. 2000).

In this paper we combine our spherical harmonic decomposition,
parameter eigenmode analysis and hierarchical data compression
methods to analyse the PSCz. We study both nonlinear
multiparameter estimation, the redshift-space distortion
parameter and the amplitude of power, using hierarchical 
data compression, from the
harmonic modes of the PSCz survey. The increase in analysing power
using these methods allows us to increase the number of modes
available for study, and hence a corresponding increase in
accuracy of our results.

Padmanabhan, Tegmark and Hamilton (1999) have also used the
spherical harmonic decomposition to analyse the CFA/SSRS UZC
galaxy redshift survey, while Hamilton, Tegmark \& Padmanabhan
(2000) have applied the analysis to the PSCz redshift survey. In
both cases they employ Karhunen-Lo\`{e}ve methods to estimate the 
quadratic band-power in
these surveys. In the former survey the band-power was measured
in redshift-space, while in the latter analysis of the PSCz they
measured the real-space galaxy-galaxy, galaxy-velocity and
velocity-velocity power spectra, and estimated $\beta$ from a 
least squares fit to the ratios
of the galaxy-velocity to galaxy-galaxy and velocity-velocity to galaxy-galaxy power
spectra.

The paper can be summarised as follows. In Section 2 we briefly
review the spherical harmonic decomposition,  OMA,
parameter eigenmodes and hierarchical data compression
methods. In Section 3 we perform an maximum likelihood analysis of the PSCz,
for the redshift-space distortion parameter and real-space clustering
amplitude of galaxies.
Our results are presented in Section 4, and conclusions are made
in Section 5. We begin by a brief review of our data analysis
methods.

\section{Data Analysis}

\subsection{Spherical Harmonic decomposition}

Following HT, BHT, and T99 we can
decompose the density field of galaxies in  a
redshift survey into harmonic modes
\be
    \rho_{\ell mn} = c_{\ell n} \int \! d^3\!s \,
    \rho(\s) w(s)j_{\ell}\left(k
    s\right)
    Y^{*}_{\ell m}\left(\Omega \right),
\label{transequ} \ee where $Y_{\ell m}$, are spherical harmonics,
$j_{\ell}$ are a discrete set of spherical Bessel functions,
$w(s)$ is an adjustable weighting function and $\s$ is the
redshift-space position variable. The $c_{\ell n}$ are
normalization constants and $k=k_{ln}$ are discrete wavenumbers.
Each mode was weighted with a Feldmann-Kaiser-Peacock (1994)
weight, $w(k,s)=[1+\phi(s)P(k)]^{-1}$, where $\phi(s)$ is the
(redshift-dependent) average number density of galaxies in the
survey.

The observed coefficients, $D_{\ell m n}$, can be related to the underlying linear
density modes, $\delta_{\ell m n}$, by (HT, BHT, T99)
\be
    \D \equiv \rhob - \rhoo = \Sb \W (\Phib+\beta \Vb) \,\deltab.
\ee
The transition matrices $\Sb$, $\W$, $\Phib$ and $\Vb$ correspond to the
effects of small-scale random radial velocity distortions, the angular window
function, the radial galaxy selection function and linear redshift
space distortions, respectively. The mean field, $\rhoo$, is nonzero
due to the radial selection function and angular window function. Note that
there is no matrix multiplication implied between the angular matrix,
$\W$, and the radial matrix, $\Phib+\beta \Vb$, since these are orthogonal.

The small-scale radial velocities can be accurately modelled
by multiplying the galaxy power spectrum by a Lorentzian function.
This implies that each mode is multiplied by the square-root of
a Lorentzian. Inverse Fourier transforming we find that the density
field should be convolved with the function
\be
    S(x) = \frac{2 \sqrt{2}}{\sigma_v}
    K_0\left(\frac{\sqrt{2}}{\sigma_v}x\right),
\label{eq5} \ee where $K_n$ is an $n^{th}$-order modified Bessel
function and $\sigma_v$ is the 1-d velocity dispersion. (Note
that in the analysis of T99, the 3-d velocity dispersion was used
incorrectly. Changing to the correct value had little effect.
However, as we are pushing our model to higher wavenumbers here,
it is more important to have the correct value of the velocity
dispersion.) The transition matrix, $\Sb$, is found
by a spherical harmonic transformation of  $S(x-y)$.

Two immediate advantages of this treatment are accounting for the
effects of the monopole and dipole modes. The monopole mode
contains information about the mean density of the survey. In
previous methods this can bias down the estimated power as the
mean is estimated from the survey itself and may not be the true
mean (Tadros \& Efstathiou 1996). In our treatment the monopole
mode can be removed, effectively removing this bias (T99). As the
PSCz is not all sky, some aliasing takes place at the few percent
level, and we include the effects of this leakage, through $\W$.
The dipole includes contributions from our own motion in the
redshift space distortion. Again we can mostly remove this by
ignoring the dipole mode, and accounting for aliasing from the
angular mask (T99).

The distribution of linear harmonic modes can be described by a multivariate
likelihood,
\be
    -2 \ln \eL [\D|\beta,P(k)] =  \D^t \C^{-1} \D + \ln \det \C,
\ee
where $\C = \lgl \D \D^t \rgl$ is the data covariance matrix. Details
for dealing with non-axisymmetric angular window functions in the
covariance matrix, as well as further details of the likelihood
analysis, are given in T99.

\subsection{Generalised Optimal Mode Analysis (GOMA)}
The technical advance which allows us to reduce the error bars in the
determination of $\beta$ and the power spectrum is a new optimised
form of data compression, which we call generalised optimal mode analysis.
Details of the method will appear elsewhere (Ballinger et al. 2000),
but we sketch the main ingredients here.  

The need for data compression is twofold: first, the speed of analysis
generally scales as $N^3$, where $N$ is the number of modes
considered.  These modes might be, for example, spherical transform
coefficients.  Secondly, since the covariance matrix is computed
numerically, small numerical errors can lead to a non-positive
definite matrix, which makes no physical sense; even a single negative
eigenvalue out of several thousand is fatal.  Instead of working with
the full set of spherical transform coefficients up to some maximum
wavenumber, we form orthogonal linear combinations, and use these as
the data.  GOMA consists of two parts; one is {\em parameter eigenmode
analysis}, where instead of choosing $\beta$ and $\Delta_{0.1}$ as the two
parameters to be determined, we introduce a new parameter, which runs
along the longest likelihood ridge of Figure 1, and use the data compression
methods of TTH to
make the likelihood as narrow as possible in this direction.  Since
the length of this controls the marginal errors in {\em both} $\beta$
and $\Delta_{0.1}$, the method is very effective. 
It is worth noting that only OMA will determine the best set of eigenmodes
following this procedure, since we now have a linear combinations of
parameters which are nonlinear in the covariance matrix.

The second (optional) ingredient is {\em hierarchical data compression}.  We
cannot find the optimal linear combinations of the entire mode set,
because of the numerical problems identified above.  Instead, we form
optimal orthogonal linear combinations of subsets of the modes (in
discrete wavenumber ranges), and then combine the best modes to form a
new set.  This process can be repeated hierarchically to produce a
near-optimal set of modes.   Note that when mode sets are combined
they are not orthogonal and we use the correct correlation properties.

\section{Application to the PSCz}

We have applied our methods to the new IRAS PSCz redshift survey (Saunders
et al 1999), with a flux cut of $0.75$Jy, and the ultra-conservative mask
created by T99, corresponding to optical extinctions of $A_B=0.75$mag and
excluding $35\%$ of the sky.
Our flux cut is above the formal limit of of $0.6$Jy for the catalogue,
but motivated by a systematic effect we found linked to flux in T99.
This appeared as a flux-dependency on the amplitude of real space
perturbations, where the low-flux galaxies had roughly a factor of
two higher clustering amplitude, the cause of which we were unable to identify.
However by cutting the catalogue at $0.75$Jy the effect could be removed.
This cut in flux, along with the more conservative mask left us with
around 7000 galaxies.

The survey was put in a sphere of radius $r_{\rm max}=200 \hMpc$ and
transformed into spherical harmonic modes. We analysed all the modes down
to $k=0.2 h {\rm Mpc}^{-1}$, with $n = 1-14 $ and $\ell=2-36$
yielding a total of $4644$ harmonic modes. Modes in the range $n=0-20$ and
$\ell=0-50$ were used for the convolution matrices. 
Since we use a higher cut in wavenumber, $k$, than in the
analysis of T99 we must be cautious about introducing nonlinear
modes which go beyond our analysis. The main concern is the effect of
``fingers-of-god'' contaminating our analysis. The effect of these would
be to lower the measured value of $\beta$, since its effect is to
elongate structure along the line of sight, and lower the
clustering amplitude. We have tested our
methods using CDM simulations under similar conditions, and find that
our correction for radial small-scale velocities, equation (\ref{eq5}),
is accurate (Ballinger et al 2000).

The $4664$ modes were compressed down to $2278$, using the
hierarchical compression method once, and a new likelihood
analysis applied. In T99 around $1300$ modes where analysed. In this analysis
we have partly made use of the increase in computing power in the time
between the two analyses. 
However the real limiting feature of our previous analysis was 
numerical instability problems in the covariance matrix. Small errors can
produce a covariance matrix which is not positive-definite, which 
makes no physical sense (a probability distribution in data space
which grows exponentially along one principal direction). 
A great advantage of the compression
mechanism is that this numerical problem is completely solved:
the best modes for parameter estimation are well-behaved, and the
covariance matrix inversion proceeds smoothly.
Hence the overall time for computing our results (around 1 weeks CPU time) 
remained constant. 

\section{Results}

\subsection{Likelihood Analysis}

\begin{figure}
\centering
\begin{picture}(200,180)
\includegraphics{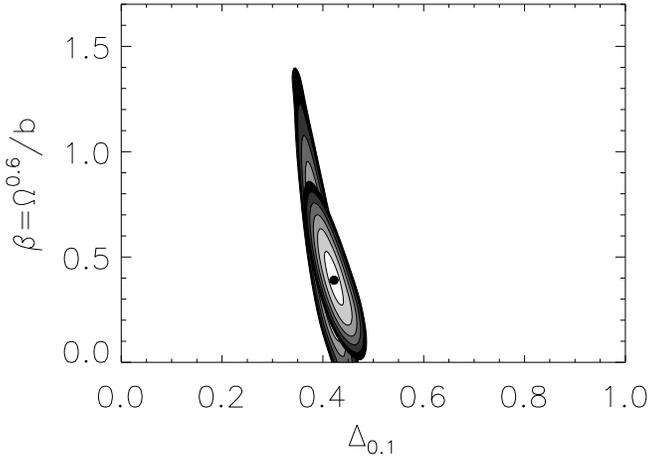}
\end{picture}
\caption{Likelihood contours for the parameter space of redshift
space distortions, $\beta$, and real-space galaxy power, $\Delta_{0.1}$,
measured at $k=0.1 h {\rm Mpc}^{-1}$ for the PSCz survey cut at a
flux limit of $0.75$Jy. The larger set of contours is the results of the
Tadros et al (1999) analysis using $1300$ modes and $k_{\rm max}=0.13 \Mpch$.
The smaller set of contours is the present analysis using $4644$ modes
compressed into $2278$ modes and $k_{\rm max}=0.20 \Mpch$. In both analyses
a conservative mask was used leaving a total sample of $7042$ galaxies.
The contours are spaced by $\Delta \ln \eL=-0.5$.}
\label{fig1}
\end{figure}

Figure \ref{fig1} shows the results of our likelihood analysis for
the $\beta - \Delta_{0.1}$ plane. The larger contours are the
results of T99 for 1300 modes. Contours are plotted at intervals
of $-0.5$ in $\ln \eL$ from the maximum. The smaller set of
contours are for our new analysis for $4644$ harmonic modes
compressed to $2278$. We have used a CDM-type power spectrum with
shape parameter $\Gamma=0.2$ to calculate the covariance
matrices, leaving the amplitude a free parameter, and in the
wavenumber-dependent mode weighting function. We used a value of
$\sigma_v=224 \kms$ for the 1-dimensional velocity dispersion of
galaxies in the scattering matrices, but have also experimented with $112 \kms$
and $336 \kms$.

The increase in parameter information yields a new, lower value of
\be
    \beta = 0.39 \pm 0.12,
\ee
where we quote the marginalized errors. This is
consistent with our previous result, but with a much smaller error
ellipse, around a factor of 3 improvement, which significantly rules
out both $\beta=1$ and $\beta=0$. The amplitude of the real space
galaxy power spectrum is
\be
    \Delta_{0.1} = 0.42 \pm 0.02
\ee which is again consistent with the results of T99 with a
slightly smaller error. The covariance parameter of $\beta$ and
$\Delta_{0.1}$ is $r=0.82$, estimated from the ellipticity of the
error ellipse. We find no evidence that our analysis is being
biased by nonlinear effects, since then we would expect a sudden
change in both the value of $\beta$ and a drop in the amplitude
of real-space perturbations, neither of which is seen.  We also
find that the maximum likelihood values are almost unchanged, moving by 
less than $\Delta \beta \approx 0.1$, if
we change the velocity dispersion to $112$ or $336 \kms$,
suggesting that nonlinear effects are not significant for this
analysis.

As the amplitude of galaxy perturbations is proportional to the bias factor,
we can combine $\beta$ and the linear galaxy power spectrum to estimate the
amplitude of the mass power spectrum,
\be
    \Delta_m (k) = \beta \Delta_{\rm gal} (k) \Omega_m^{-0.6}.
\ee
The real-space power spectrum of the PSCz was estimated by T99, and
is plotted in Figure 2. Also shown is the amplitude of mass perturbations
for $\Omega_m=0.16$. The point at $k=0.23 \Mpch$ is the mass amplitude
on a scale of $8\hMpc$, estimated from the abundance of clusters (Henry \&
Arnaud 1991, White, Efstathiou \& Frenk 1993, Viana \& Liddle 1996, Eke,
Cole \& Frenk 1996),
\be
    \sigma_8 = 0.56 \Omega_m^{-0.47},
\ee
with a conservative error of around $30\%$. The mass amplitude
from the PSCz at $k=0.1 \Mpch$ is
\be
    \Delta_m (0.1) = (0.16 \pm 0.04) \Omega_m^{-0.6}.
\ee

In Figure 2 we plot the mass amplitudes  for the PSCz and the
cluster abundance. There is a clear consistency between these two
estimates of the mass amplitudes. Assuming bias is linear on
large scales, estimating mass amplitudes from large-scale redshift
surveys is simpler, and hence in principle more robust, than the
abundance of clusters argument. In addition this can be used to
sample the linear spectrum of mass perturbations on a range of
scales, rather than being restricted to one scale.

\begin{figure}
\centering
\begin{picture}(200,200)
\includegraphics{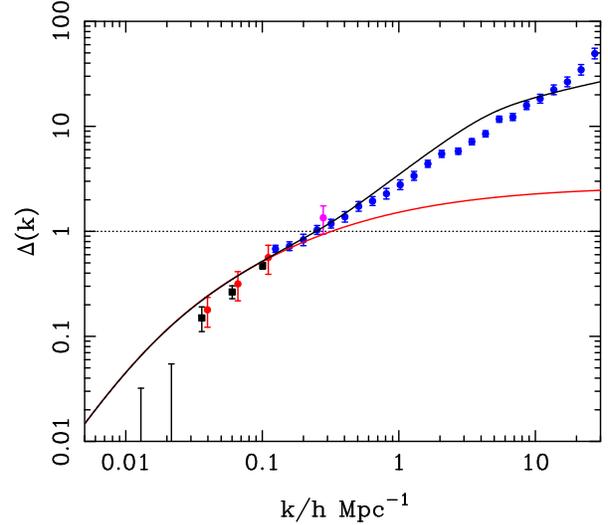}
\end{picture}
\caption{The real-space power spectrum of IRAS galaxies. On large, linear
scales the analysis of Tadros et al (1999) is used to estimate the
real-space band-power in five passbands. On smaller scale we have plotted
the real-space power spectrum estimated from the analysis of Saunders et al
(1997) from the cross-correlation of the QDOT and QIGC surveys. The
lighter points are the mass amplitudes estimated from the redshift space
distortion parameter. The last point is the mass amplitude estimated
from the abundance of clusters. The model
 fit is a $\Lambda$CDM model with $\Omega_\Lambda=0.84$ and
$\Omega_m=0.16$, $h=0.65$. The lighter line is the linear fit, while the
solid line uses the nonlinear transformation of Peacock \& Dodds (1996).}
\label{fig2}
\end{figure}

\subsection{Cosmological Parameters}

Although the shape of the real-space, linear galaxy power spectrum can
be used to constrain models of structure formation, the range of points
in the linear regime is limited. A better test of models is to compare
the mass amplitudes from the PSCz against the amplitude estimated
from the CMB. To span the scales between different measurements
we need to assume a model linear mass power spectrum. We shall assume
that a standard CDM-type power spectrum of the form
\be
    \Delta^2_m(k) = Q^2 (k/H_0)^4 T^2(k,\Omega_m,h),
\ee
where we use the CDM transfer function of Bond \& Efstathiou (1984), which 
gives a reasonable fit for a range of CDM models.
This provides us with a 3-parameter model dependent on the parameter set
$\Omega_m$, $h$ and $Q$\footnote{Our parameter Q is equivalent to the parameter
$\delta_H$, the amplitude of clustering at the present Hubble scale, used elsewhere.}. 
This is an interesting parameter set, in
particular since CMB data alone cannot constrain $\Omega_m$ without
the addition of LSS data.

If we leave $h$ as a free parameter, the fit
diverges off to high $h$ and $Q$ and low $\Omega_m$, preserving the shape, but 
increasing the clustering amplitude. To avoid this we
impose the HST constraint that $h=0.65\pm0.12$. This effectively removes
a degree of freedom in the fit. Beyond this we fit the PSCz
mass points, the abundance of clusters constraint and the COBE 4-yr
normalisation for a flat universe (Bunn \& White 1997);
\be
    Q =  (1.94 \pm 0.2)\times 10^{-5} \Omega_m^{-0.79}.
\ee 
We implicitly assume that the universe is spatially flat, as
implied by the recent Boomerang (Lange et al 2000) and MAXIMA
(Hanany et al. 2000, Balbi et al. 2000) results and that the
spectral index is unity, again as implied by CMB results. We also
implicitly assume that the contribution to the mass density from
baryons and neutrinos is negligible.

\begin{figure}
\centering
\begin{picture}(200,200)
\includegraphics{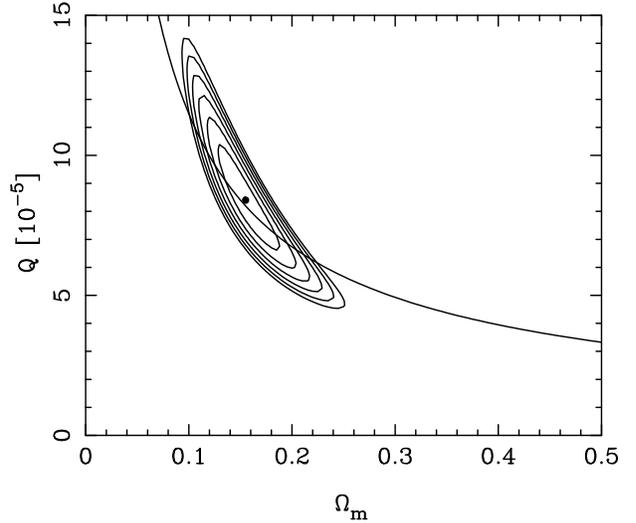}
\end{picture}
\caption{The $\chi^2$ fit to the linear real-space PSCz mass
power spectrum, the abundance of clusters
and the CMB for the two parameter space of mass-density
parameter, $\Omega_m$, and the amplitude of mass perturbations, $Q$.
The solid line is the constraint from the 4-year COBE analysis of the
CMB.}
\label{fig3}
\end{figure}

Figure 3 shows the $\chi^2$ distribution for the $\Omega_m-Q$ plane,
with $h=0.65$. The minimum $\chi^2$ values are
\be
    \Omega_m = 0.16 \pm 0.03 , \hspace{1.cm} Q = (8.4 \pm 3.8) \times 10^{-5}
\ee 
where we quote marginal errors. We can also obtain an estimate of the 
bias parameter for IRAS galaxies:
\be
	b = 0.84 \pm 0.28,
\ee
and the spectral shape parameter,
\be
	\Gamma = 0.1 \pm 0.03,
\ee
which is somewhat lower than the usual value of $0.2$ from the shape of the galaxy 
clustering spectrum alone.
This best fit has a
$\chi^2=2$ for 3 degrees of freedom, and so lies within the range
of acceptable fits. However, while suggestive, this analysis
makes more assumptions than one would like. These assumptions can
be dropped by a combined analysis of the recent Boomerang and
MAXIMA results of the small-angle fluctuations in the CMB, which
goes beyond our present analysis.

In Figure 2 we plot the PSCz real-space galaxy power spectrum and
real-space mass power spectrum, along with our best-fit model, a
$\Lambda$CDM, normalised to the COBE 4-year data. The model fits
over two orders of magnitude in scale, but nonlinear and nonlocal
bias effects lead to a disagreement on smaller scales, below
$k=0.3 \Mpch$. Recent results from Seljak (2000), Ma \& Fry
(2000), and Peacock \& Smith (2000) suggest that this nonlinear
regime can best be understood in terms of a superposition of
randomly positioned collapsed clusters, integrated over the
cluster mass function. Seljak (2000), and Peacock and Smith (2000)
go further to argue that the nonlinear bias function of galaxies
arises due to the statistics of the occupation number of galaxies
in haloes and show agreement between the $\Lambda$CDM model and
the small-scale power-law spectrum of galaxies. Perhaps the
analysis of this regime is not as daunting as it once appeared.

\section{Conclusions}

In this paper we have presented a new analysis of the PSCz
redshift survey. Using the spherical harmonic analysis of
HT, BHT and T99 to decompose the survey, taking into
account the effects of linear redshift-space distortions,
nonlinear ``fingers-of-god'' effects, limited sky coverage,
and the radial distribution of galaxies in the survey, we
have applied a likelihood analysis to a conservative cut of the
PSCz survey. The catalogue was cut at $0.75$Jy and a
conservative mask used to avoid systematic uncertainties at the
low-flux end of the catalogue and near the galactic plane.
The spherical harmonic analysis of the remaining $7042$ galaxies
resulted in $4644$ harmonic modes. We used a hierarchical
data compression to reduce this to $2278$ for the final analysis.
The compression
was applied using a parameter eigenmode approach that compresses
information along the line of largest degeneracy in parameter
space, thus ``squeezing'' the uncertainty in this direction.

We used these methods to estimate the redshift-space distortion parameter,
$\beta$ and the amplitude of the real-space galaxy power spectrum,
parameterised by the amplitude, $\Delta_{0.1}$, at a wavenumber $k=0.1 \Mpch$.
Applying the likelihood analysis to wavenumbers below $k=0.2 \Mpch$, where
linear theory will hold and ``fingers-of-god'' effects can be corrected
for, we find
\ba
        \beta &=& 0.39 \pm 0.12, \\
        \Delta_{0.1} &=& 0.42 \pm 0.02
\ea
quoting marginalised uncertainties. The distortion parameter is
slightly lower than the uncompressed analysis of Tadros et al, with
an uncertainty reduced by over a factor of 2. The
consistency of these results with that of the
earlier analysis of T99 leads us to believe that
our analysis has not been heavily contaminated by nonlinear effects,
while the conservative cuts in flux and sky coverage avoid
contamination by flux systematics. These results also are in agreement 
with other, independent determinations of the distortion parameter.
Comparing the velocity field reconstructed from the PSCz with the ENEAR 
survey, Nusser at al (2000) find $\beta=0.5 \pm 0.1$, while Valentine,
Saunders \& Taylor (2000) find $\beta =0.5 \pm 0.1$ by comparing the 
reconstructed dipole with the observed dipole, the bulk flow of galaxies 
with the MKII survey and the dipole again with the SFI catalogue.
Finally, using a least squares fit to the ratios
of the galaxy-velocity to galaxy-galaxy and velocity-velocity to galaxy-galaxy power
spectra, Hamilton, Tegmark \&  Padmanabhan find $\beta =0.41 \pm 0.13$ for the
PSCz.

Since in the linear regime the
galaxy power spectrum is likely to be proportional to the mass power
spectrum we can combine $\Delta_{0.1}$ and $\beta$ to find the amplitude
of the mass perturbations,
\be
    \Delta_m (0.1) = (0.16 \pm 0.04) \Omega_m^{-0.6}.
\ee 
Combined with the constraints from the CMB, and HST
observations of the Hubble parameter, and assuming CDM, a flat
universe, scale invariant initial mass perturbations and
negligible contribution to the total mass-density from baryons
and neutrinos we find $\Omega_m =0.16 \pm 0.03$, and a bias
parameter for IRAS galaxies of $b=0.84 \pm 0.28$. The minimum
$\chi^2$ fit has a value of $\chi^2 =2$ for 3 degrees of freedom.

\section*{Acknowledgements}

We thanks the PSCz team, and especially Will Saunders, for help in
understanding the intricacies of the PSCz catalogue. ANT, WEB and
HT thank the PPARC for postdoctoral support.  Computations were
made using STARLINK facilities.

\section*{References}

\bib Balbi A., et al ., 2000, astro-ph/0005124

\bib Ballinger W., 1997, Phd Thesis, Univ. Edinburgh

\bib Ballinger W., Heavens A.F., Taylor A.N., 1995, MNRAS, 276, 59p

\bib Ballinger W., Taylor A.N., Heavens A.F., Tadros H., 2000, in
preparation

\bib Bond, J.R., 1995, Phys. Rev. Lett., 74, 4369

\bib Bond, J.R., Efstathiou G., 1984, ApJ, 285, L45

\bib Bunn E.F., White M., 1997, ApJ, 480, 6

\bib Eke V.R., Cole S.,  Frenk C.S., 1996, MNRAS, 282, 263

\bib Feldmann H.A., Kaiser N., Peacock J.A., 1994, ApJ, 426, 23

\bib Hanany et al., 2000, astro-ph/0005123

\bib Hamilton A.J.S., Tegmark M., Padmanabhan N., 2000, submitted to MNRAS
    (astro-ph/0004334)

\bib Heavens A.F., Taylor A.N., 1995, MNRAS, 275, 483

\bib Henry J.P., Arnaud K.A., 1991, ApJ, 372, 410

\bib Lange A.E., et al., 2000, astro-ph/0005004

\bib Ma C.-P., Fry J.N., 2000, astro-ph/0003343

\bib Nusser A., et al, 2000, astro-ph/0006062

\bib Padmanabhan N., Tegmark M., Hamilton A., 1999, submitted to ApJ
     (astro-ph/9911421)

\bib Peacock J.A., Dodds S., 1996, MNRAS, 280, L19

\bib Peacock J.A, Smith R.E, 2000, astro-ph/0005010

\bib Peebles P.J.E., 1980, ``Large-Scale Structure of the Universe'',
    Princeton Univ. Press, Princeton, NJ

\bib Saunders W., Rowan-Robinson M., Lawrence A., 1997

\bib Saunders W., et al., 2000, astro-ph/0001117

\bib Seljak U., 2000, astro-ph/0001493

\bib Tadros H., Efstathiou G.P.E., 1996, MNRAS, 282, 1381

\bib Tadros H., et al., 1999, MNRAS, 305, 527

\bib Taylor et al., 1998, in proc Cambridge Particle Physics and
    Early Universe Conf. (astro-ph/9707265)

\bib Tegmark M., Taylor A.N.,  Heavens A.F., 1997, ApJ, 480, 22

\bib Valentine H., Saunders W., Taylor A.N., 2000, submitted to MNRAS 
	(astro-ph/0006040)

\bib Viana P.T., Liddle A.R., 1996, MNRAS, 281, 323

\bib White S.D.M., Efstathiou G.P., Frenk C.S., 1993, MNRAS, 262, 1023

\end{document}